\documentclass[dvips]{article}
\usepackage{icrctc07}
\title{Analytical description of the Day-Night neutrino asymmetry}
\shorttitle{Day-Night neutrino asymmetry}
\authors{A. D. Supanitsky$^1$, J. C. D'Olivo$^1$, and G. A. Medina-Tanco$^{1}$.}
\shortauthors{A. D. Supanitsky and et al}
\afiliations{$^1$Departamento de F\'isica de Altas Energ\'ias, Instituto de Ciencias Nucleares, Universidad Nacional Aut\'onoma de M\'exico, A. P. 70-543, 04510, M\'exico, D. F., M\'exico.}
\email{dalesupa@gmail.com}

\abstract{ We present a new treatment of the Earth matter effects on neutrino oscillations
that is valid for an arbitrary density profile. When applied to the study of the day-night
effect on the solar neutrino flux it renders a simple analytical expression, which is more
accurate than those derived by using the perturbation theory and can be extended to higher energies.}

\begin{document}
\maketitle
%Begin the section.

\section{Introduction}

Different types of experiments have provided compelling evidence for
neutrino oscillations \cite{Smirnov}. In the case of solar neutrinos the leading effects
can be accounted by oscillations between two neutrino flavors, parameterized
in terms of the mass square difference $\delta m^2=m_2^2-m_1^2$ and the mixing
angle $\theta$. A global fit of all the existing data gives
$\delta m^2=(7.9-8)\times10^{-5}$ eV$^2$ and $\sin^2\theta=0.310-0.315 \,$\cite{Smirnov},
which is in good agreement with the results of other groups. These values
belong to the region in the parameter space referred to as the Large Mixing
Angle Solution (LMA). According to the LMA, the $^8$B electron neutrinos produced
in the Sun undergo a highly adiabatic conversion and are almost totally converted
into the mass eigenstate $\nu_2$. Then, the electron neutrino survival probability
is $P(\nu_e \rightarrow\nu_e) \cong \sin^2 \theta$. However, during the night solar
neutrinos arriving to terrestrial detectors  travel a certain distance through the
Earth's matter, which affects the oscillations pattern. This leads to a partial regeneration
of the electron neutrino flux, a phenomenon known as the day-night effect.

Matter effects on the neutrino oscillations inside the Earth are conveniently
taken into account  in terms of the parameter
$\varepsilon(t) \equiv {2E V(t)}/{\delta m^2}$,
where $V(t)=\sqrt{2} G_F n_e(t)$ represents the potential energy for $\nu_e$,
which comes from the charged-current interaction with electrons. Here, $G_F$ is
the Fermi constant, $E$ is the neutrino energy, and $n_e(t)$ is the number density 
of electrons along the neutrino path. In terms of the Avogadro number $N_A$,
\begin{eqnarray}
\varepsilon(t)\!\!\!&\cong&\!\!\!0.019 \!\left[\frac{E}{10 \textrm{ MeV}}
\right]\!\!\! \left[\frac{n_e(t)}{N_A \ \textrm{cm}^{-3}}\right]
\nonumber \\
&&\times \left[\frac{8\times10^{-5} \textrm{
eV}^2}{\delta
m^2}\right], \label{Epsilon}
\end{eqnarray}
For the favored value of $\delta m^2$ and the energy range of solar neutrinos,
Earth's density is such that $\varepsilon \ll1$.  Taking advantage of this fact, 
perturbation theory has been applied to derive an analytical expression for the
day-night rate asymmetry to first order in $\varepsilon$ \cite{Valle:04,Ioannisian:04},
which is valid for any density profile. The method simplifies the numerical
calculations  and it has been subsequently improved by means
of a  second order expansion in $\varepsilon$ \cite{Ioannisian:05}. In this work we show
that  a convenient alternative to the perturbative approach is provided by the Magnus 
expansion of the evolution operator \cite{JCDOlivo:92} and from it we derive a 
more accurate formula for the regeneration probability.

\section{Neutrino Oscillations in Matter}

We consider a system consisting of two neutrino flavors,
$\Psi_f=(\Psi_e, \Psi_\mu)$, which are related to the mass
eigenstate, $\Psi_{mass}=(\Psi_1, \Psi_2)$, according to

\begin{equation}
\label{Bases}
\Psi_{f}=U(\theta)\Psi_{mass},
\end{equation}
where,
\begin{equation}
\label{U}
U(\theta)=\left(%
\begin{array}{cc}
  \cos\theta & \sin\theta \\
  -\sin\theta & \cos\theta \\
\end{array}%
\right).
\end{equation}

The evolution operator of the system satisfies the equation
\begin{equation}
\label{TimeEvol}
i\frac{d\mathcal{U}}{dt}(t,t_{0})=H(t)\ \mathcal{U}(t,t_{0})\,,
\end{equation}
with the initial condition $\mathcal{U}(t_{0},t_{0})=1\!\!1$.
The hamiltonian in the mass base is given by
\begin{eqnarray}
\label{Hamiltonian}
H(t)\!\!\!&=&\!\!\!\left(%
\begin{array}{cc}
  0 & 0 \\
  0 & \frac{\delta m^2}{2 E} \\
\end{array}%
\right)+ \nonumber \\
&& \!\!\!V(t)\!\!\left(%
\begin{array}{cc}
  \cos^2\theta & \sin\theta \cos\theta \\
  \sin\theta \cos\theta & \sin^2\theta \\
\end{array}%
\right)\!,
\end{eqnarray}
and its eigenvalues are
\begin{equation}
\label{Eigen}
\lambda_{\pm}(t)=\frac{1}{2}[V(t)+\frac{\delta m^2}{2 E}\pm
\Delta_{m}(t))]\,, 
\end{equation}
with
\begin{equation}
\label{Delta}
\Delta_{m}(t)=\frac{\delta m^2}{2 E}
\sqrt{(\varepsilon(t)-\cos 2\theta)^2+\sin^2 2\theta}\,.
\end{equation}

Let us now write
\begin{eqnarray}
\label{Upop}
\mathcal{U}(t,t_0)&=&\mathcal{P}(t,t_0)\
\mathcal{U_{P}}(t,t_0), \\
\mathcal{P}(t,t_{0})&=&\left(%
\begin{array}{cc}
  e^{-i \alpha_{-}(t,t_0)}& 0 \\
  0 & e^{-i \alpha_{+}(t,t_0)} \\
\end{array}%
\right),\nonumber
\end{eqnarray}
where
$
\alpha_{\pm}(t,t_0)\!\!=\!\!\int_{t_0}^t dt' \lambda_{\pm}(t').
$
The operator $\mathcal{U_{P}}(t,t_0)$ obeys Eq. (\ref{TimeEvol})
but for the Hamiltonian $H_{\mathcal{P}}(t,t_0)=\mathcal{P}^\dag(t,t_0)[H(t)-H_D(t)]
\mathcal{P}(t,t_0)$, where
$H_D(t)=diag(\lambda_{-}(t),\lambda_{+}(t))$. By expanding $\lambda_{\mp}$
to first order in $\varepsilon(t)$ we obtain an approximated
expression for $H_{\mathcal{P}}$ with vanishing elements in the diagonal:
\begin{equation}
\label{HamPapp}
H_{\mathcal{P}}(t,t_0)\!\cong \!V(t)\frac{\sin2\theta}{2} \left(%
\begin{array}{cc}
  0 & e^{-i \phi_{t_0\rightarrow t}} \\
  e^{i \phi_{t_0\rightarrow t}} & 0 \\
\end{array}%
\right),
\end{equation}
with
$
\phi_{t_0\rightarrow t}=\int_{t_0}^t dt' \Delta_m(t').
$

The relevant quantity is the regeneration probability defined as the
difference between the day and night probabilities,  $F_{reg}(E)\equiv P_{2
\rightarrow e}(E)-\sin^2\theta$, where $P_{2 \rightarrow e}(E)=|
\langle \nu_e|\hat{\mathcal{U}}(t,t_0)|\nu_2 \rangle |^2$.
Here, we determine the evolution operator in the mass base from
Eq. (\ref {Upop}) by evaluating $\mathcal{U_{P}}$
in terms of the lowest-order Magnus approximation,
$\mathcal{U_{P}}(t,t_0) \cong \exp[-i \int_{t_0}^t dt' H_{\mathcal{P}}(t',t_0)]$.
Proceeding in such a way we get
\begin{eqnarray}
\label{FregSim}
F_{reg}(E) &=& \frac{1}{2} \sin (2I) \sin 2\theta
\sin(\phi_{\bar{t}\rightarrow t}) \nonumber \\
&& +\sin^2 (I) \cos 2\theta,
\end{eqnarray}
with
\begin{equation}
\label{Idef}
I=\sin 2\theta \int_{\bar{t}}^t dt' V(t')
\cos(\phi_{\bar{t}\rightarrow t'}).
\end{equation}
In writing Eq. (\ref{FregSim}), we assumed that the potential is symmetric with respect to
the middle point of the trajectory $\bar{t}=(t+t_0)/2$, which is the situation for a medium like the 
Earth, with a spherically symmetric density profile.
By keeping the lowest order terms of the expansion in $I$, our result for
$F_{reg}(E)$ reduces to the one calculated to first order in $\varepsilon$
\cite{Ioannisian:05}. 

In order to make a numerical comparison of the different formulas, we examine
the case of a neutrino that crosses the Earth passing trough its center. For the 
electron density we adopt the simplified model called mantle-core-mantle
\cite{Stacey:77}.  According to it, $n_e(r)$ is approximated by a
step function and the radius of the core and the thickness of the mantle are
assumed to be half of the Earth radius.  Accordingly, we put
\begin{equation}
\label{V}
n_e(r)=N_A \left\{ \begin{array}{ll}
                   5.953 \textrm{ cm}^{-3}, &  r \leq R_\oplus/2 \\
                                            &                  \\
                    2.48 \textrm{ cm}^{-3}, &  R_\oplus/2 < r \leq R_\oplus
              \end{array}
    \right.,
\end{equation}
where $R_\oplus$ is the radius of the Earth.

Following Ref. \cite{Ioannisian:05}, we introduce the function
\begin{equation}
\label{delta}
\delta(E)= \frac{1}{\bar{F}_{reg}(E)}
[F_{reg}^{(appr)}(E)-F_{reg}^{(exact)}(E)],
\end{equation}
where $F_{reg}^{(appr)}$ is given by a certain (approximated)
analytical expression, $F_{reg}^{(exact)}$ is obtained from the exact
(numerical) solution, and
\begin{equation}
\label{Fbar}
\bar{F}_{reg}(E)=\frac{1}{2} \varepsilon(t_s) \sin^2 \theta
\end{equation}
is the average regeneration factor evaluated at
the surface layer. Essentially, $\delta$ represents the relative error
of the approximated expression. 

\begin{figure}
\begin{center}
\includegraphics [width=0.48\textwidth]{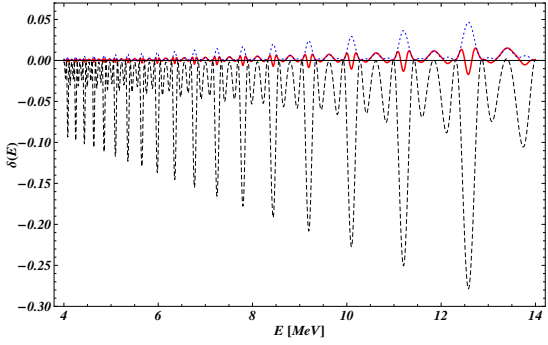}
\end{center}
\caption{Relative error $\delta$ vs the neutrino energy in the case of 
a neutrino that goes through the Earth passing by its center, for  
$\delta m^2 = 8\times 10^{-5} \textrm{ eV}^2$ and
$\tan^2\theta=0.4$.  The dashed line and dotted blue line
are the first and second order approximations in $\varepsilon$, respectively, 
and the solid red line corresponds to the first-order Magnus result.}
\label{DeltaVsE}
\end{figure}

Figure \ref{DeltaVsE} shows
$\delta$ as a function of the neutrino energy for a neutrino that 
propagates inside the Earth and goes  through its center. $F_{reg}^{(appr)}$
has been computed to first and second order in $V$ and by means 
of the result given in Eq. (\ref{FregSim}).
From the figure we see that the relative error for the Magnus approximation is always 
smaller than those corresponding to the perturbative calculations. Although 
it increases with energy it remains smaller than $\sim 2\%$ for the
largest energies of the solar neutrinos.

\section{Day-Night asymmetry}

\begin{figure}
\begin{center}
\includegraphics [width=0.48\textwidth]{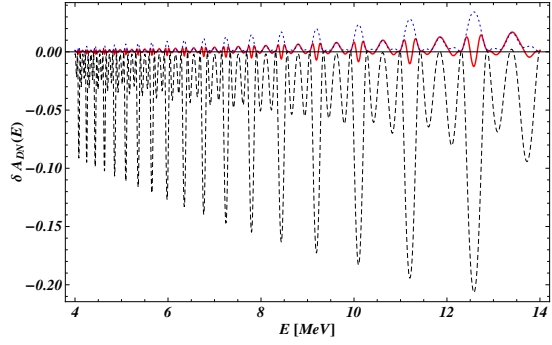}
\end{center}
\caption{Relative error in the Day-Night asymmetry as a function of the 
neutrino energy for a neutrino that propagates inside the
Earth crossing through the center ($\eta=0^\circ$). The curves correspond 
to the first (dashed line) and second (dotted blue line) order in $\varepsilon$ 
and to the Mangus result (solid red line), for $\delta m^2 = 8\times 10^{-5} \textrm{ eV}^2$ and
$\tan^2\theta=0.4$.} 
\label{DAndVsEv}
\end{figure}

\begin{figure}
\begin{center}
\includegraphics [width=0.48\textwidth]{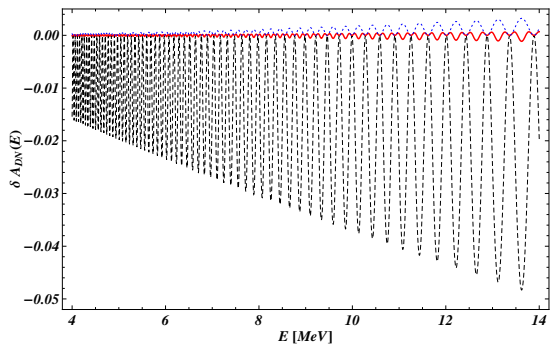}
\end{center}
\caption{Relative error in the Day-Night asymmetry as a function of the 
neutrino energy for a neutrino that propagates inside the
Earth passing tangent to the core region ($\eta=30^\circ$). The curves correspond 
to the first (dashed line) and second (dotted blue line) order in $\varepsilon$ 
and to the Mangus result (solid red line), for $\delta m^2 = 8\times 10^{-5} \textrm{ eV}^2$ and
$\tan^2\theta=0.4$.} 
\label{DAndVsEo}
\end{figure}

As a function of the energy the day-night asymmetry can be
expressed as\cite{Valle:04}
\begin{equation}\label{ADNE}
A_{DN}(E)=\frac{2\ \langle \cos 2\hat{\theta} \rangle F_{reg}}%
{1-\langle \cos 2\hat{\theta} \rangle (F_{reg}-\cos2\theta)},
\end{equation}
where,
\begin{eqnarray}
\langle \cos 2\hat{\theta} \rangle(E)\!\!\!\!&=&\!\!\!\!\!\int_0^{R_{\odot}} dr f(r)\nonumber \\
&&\!\!\!\!\!\times \frac{\cos2\theta-\varepsilon(E,r)}%
{\sqrt{(\varepsilon(E,r)-\cos2\theta)^2+\sin^22\theta}}. \nonumber \\
&&
\label{MCos}
\end{eqnarray}
Here, $f(r)$ is the spatial distribution function of the solar neutrino sources
\cite{Bahcall} and $\varepsilon(E,r)$ is determined by Eq. (\ref{Epsilon}) with 
$n_e(r)$ now representing the electron density within the Sun \cite{Bahcall}. 
Figures \ref{DAndVsEv} and \ref{DAndVsEo}
show the relative error in $A_{DN}(E)$ as a function of the energy for the three
approximations examined here and a neutrino trajectory with nadir angle $\eta=0^\circ$
(neutrino passing through the Earth center) and $\eta=30^\circ$ (neutrino passing tangent
to the core region), respectively.  We used the function $f(r)$ corresponding to the $^8$B
neutrinos and in both cases the smallest relative error is obtained with our
expression for the regeneration probability. We also see that for all the approximations
the relative error is smaller for $\eta=30^\circ$, which is due to the fact that the electron
density, and therefore $\varepsilon$, is smaller in the mantle region of the Earth.

\begin{figure}
\begin{center}
\includegraphics [width=0.48\textwidth]{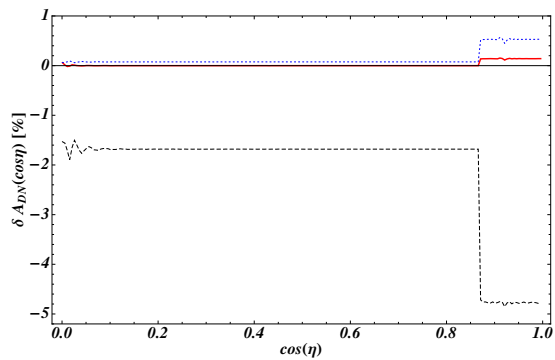}
\end{center}
\caption{Relative error in the integrated Day-Night asymmetry
as a function of the nadir angle. The curves correspond 
to the first (dashed line) and second (dotted blue line) order in $\varepsilon$ 
and to the Mangus result (solid red line), for $\delta m^2 = 8\times 10^{-5} \textrm{ eV}^2$ and
$\tan^2\theta=0.4$.}
\label{DAndVsta}
\end{figure}

Finally, we also calculate the integrated day-night asymmetry,
\begin{eqnarray}
\label{IntAdn}
A_{DN}\!\!\! &=&\!\!\!2 \int_{E_{th}}^\infty dE \phi_\nu(E) \langle \cos 2\hat{\theta} \rangle(E) F_{reg}(E) \nonumber \\
&& \times \Big[1-\int_{E_{th}}^\infty dE \phi_\nu(E) \langle \cos 2\hat{\theta} \rangle(E) \nonumber \\
&& \times \Big(F_{reg}(E)-\cos2\theta\Big)\Big]^{-1},
\end{eqnarray}
where $\phi_\nu(E)$ is the normalized flux of solar $^8$B neutrinos and $E_{th}=5$ MeV is
the detection energy threshold for Super-Kamiokande and SNO. Figure \ref{DAndVsta}
shows the relative error in $A_{DN}$ as a function of the cosine of
the nadir angle for the three approximated formulas. It can be seen that there are two
regions: one corresponding to the propagation in the mantle, $0<\cos\eta<\sqrt{3}/2$,
and the other to the propagation in the mantle and the core, $\cos\eta>\sqrt{3}/2$.
The relative error is practically constant in both regions. In the mantle it takes the values
$-1.7\%$, $0.07\%$, and $-0.001\%$ for the first order in $\varepsilon$, the second order
in $\varepsilon$, and formula (\ref {FregSim}), respectively. In the core-mantle the corresponding values are
$-4.7\%$, $0.51\%$, and $0.13\%$.

\section{Conclusions}

In this work we have applied the Magnus expansion of the time evolution operator to find approximated
analytical solutions of the system of two neutrino flavors coupled very weakly with matter.  From this 
result we derived new expressions for the regeneration probability and the Day-Night asymmetry which 
give better approximations to the exact numerical results than those obtained by using a perturbative 
approach.

\section{Acknowledgements}
This work has been partially suported by CONACYT Grant 46999-F and by PAPIIT-UNAM
Grants IN115607 and IN115707.

%This is the reference to .bib file (Whitout .bib!)
\bibliography{icrc1190}
%This in the bibtex style, is ok.
\bibliographystyle{plain}

\end{document}